\documentstyle[12pt,worldsci]{article}

\newcommand{\Avg}[1]{\left\langle #1 \right\rangle}

\input epsf.tex

\def\etal{{\it et.al.}\/}
\def\ie{{\it i.e.}\/}

\def\ol#1{\overline{#1}}

\def\hf{\frac{1}{2}}
\def\roughlyup#1{\mathrel{\raise.3ex\hbox{$\sim$\kern-.75em
\lower1ex\hbox{$#1$}}}}
\def\roughlydown#1{\mathrel{\raise.3ex\hbox{$#1$\kern-.75em
\lower1ex\hbox{$\sim$}}}}

\def\lsim{\roughlydown<}
\def\gsim{\roughlydown>}

\def\Avg#1{\langle #1 \rangle}

\def\scrs{\scriptscriptstyle}

\def\Sc#1{{{\cal #1}}}
\def\ss#1{{{\scriptscriptstyle #1}}}

\def\eqa{\begin{eqnarray}}
\def\eeqa{\end{eqnarray}}
\def\eq{\begin{equation}}
\def\eeq{\end{equation}}
\def\nn{\nonumber}

\def\Tr{{\rm Tr}\,}

\def\GF{G_{\scrs F}}

\begin{document}

\rightline{McGill-96/43}
\rightline{hep-ph/9611368}


\title{{\bf NEUTRINO PROPAGATION THROUGH A FLUCTUATING SUN}}
\author{C.P. BURGESS\thanks{Talk presented to Neutrino 96, Helsinki,
Finland, June 1996.}
{\it and} D. MICHAUD \\
\vspace{0.3cm}
{\em  Physics Department, McGill University \\
3600 University St.,  Montr\'eal, Qu\'ebec, Canada, H3A 2T8.}\\
}
\maketitle
\setlength{\baselineskip}{2.6ex}

{\begin{center} ABSTRACT \end{center}
{\small \hspace*{0.3cm}
We summarize a general formulation of particle propagation in fluctuating
media, as applied to the description of neutrino propagation through the sun.
It contains the familiar MSW effective hamiltonian, plus corrections which
describe neutrino interactions with fluctuations in the medium. An estimate of
the size of these corrections for a simple model of solar fluctuations is made,
with the conclusion that they can produce surprisingly large effects since they
grow with the correlation length of the fluctuation. For MSW oscillations, the
leading effect for small fluctuations is to diminish the quality of the
resonance,
making the suppression of the ${}^7$Be neutrinos an experimental probe of
fluctuations deep within the sun. Fluctuations can also provide an
energy-independent suppression factor of $\hf$, away from the resonant region,
even for small vacuum mixing angles. To be even potentially detectable, density
fluctuations must be correlated on scales of hundreds of kilometres, and have
amplitudes of a few percent or larger.
}}

\vspace{0.5cm}

The next generation of solar-neutrino detectors, including SNO and Super
Kamiokande
\cite{update}, bring us into a new era, in which neutrino solutions to the
solar-neutrino problem will be put under detailed experimental scrutiny. It
therefore behooves theorists to understand in detail the corrections to the
predictions of standard solar models, with and without neutrino oscillations.
Since standard treatments \cite{reviews} of MSW oscillations replace the solar
medium with an average electron density, one such correction consists of how
fluctuations in solar properties about this average can affect neutrino
propagation. We very briefly review here a method \cite{ourpaper} for treating
such fluctuations systematically. Limitations of space preclude our discussing
related approaches \cite{Loretietal,Rossi} to these issues.

A great deal is known about how particles like photons and neutrons propagate
through materials, and general techniques have been developed to describe this
propagation. Our approach is based on an adaptation of these ideas to neutrino
physics. In order to keep approximations explicit, we start with the general
case,
and then specialize to neutrinos within the sun. In its most general context we
consider two sectors, $A$ and $B$, where $A$ describes the degrees of freedom
we
wish to follow -- such as neutrino flavour -- and $B$ describes the other
degrees
of freedom, which we do not intend to measure -- like those of the solar
medium.
(There is a generalization of the approach for which a limited measurement to
$B$
is also performed \cite{ourpaper}.) We imagine $B$ to be essentially unchanged
by
its interaction with sector $A$. All predictions for measurements purely within
sector $A$ are described by the reduced density matrix, defined by tracing the
full density matrix over the degrees of freedom in $B$: $\rho_\ss{A}(t) \equiv
\Tr_\ss{B} \Bigl[ \rho(t) \Bigr]$.

The time evolution of $\rho_\ss{A}$ is computed using the hamiltonian $H =
H_0 + V$, where $H_0 = H_\ss{A} + H_\ss{B}$ describes the separate evolution of
sectors $A$ and $B$, and $V$ describes their mutual interaction. There are two
important timescales which govern this time evolution, and considerable
simplification arises when they are very different. The first scale, $\tau_p$,
is
defined as the time beyond which perturbation theory in $V$ fails. The second
scale, $\tau_c$, is the correlation time, defined as the time above which
$\Avg{\delta V(t) \delta V(t')}$ becomes negligible (if such a time exists).
Here
the average, $\Avg{\cdots} = {\rm Tr}_\ss{B} \, (\rho \cdots)$, is over sector
$B$,
and $\delta V \equiv V - \ol{V}$ is the deviation of the interaction
hamiltonian
from the effective interaction, $\ol{V}$, which describes the average evolution
in
sector $A$. In perturbation theory $\ol{V}(t) = \Avg{V(t)} -i \int_{t'}^t d\tau
\;
\Avg{ \delta V(t) \;  \delta V(\tau) } + O \left( V^3\right)$.

If $\tau_c \ll \tau_p$, then $\rho_\ss{A}$ satisfies a master equation which
can be
derived in perturbation theory, and then integrated to obtain its large-time
behaviour, even for $t> \tau_p$. It is given, to second order in $V$, by:
\eqa
{\partial \rho_\ss{A} \over \partial t}  &=& -i \Bigl[ \ol{V}(t) \rho_\ss{A}(t)
-
\rho_\ss{A}(t) \ol{V}^*(t) \Bigr] \\
&&  +  \int_{t'}^t d\tau \; \Tr_\ss{B} \Bigl( \delta V(t) \, \rho(t)
\, \delta V(\tau)  +  \delta V(\tau) \, \rho(t) \,  \delta V(t) \Bigr)
 + O\left(V^3\right) .\nn
\eeqa

These ideas may be applied to obtain an equation describing the evolution
of neutrino flavour within the solar medium \cite{ourpaper}. The result is a
two-by-two matrix equation:
\eq
\label{explme}
{\partial \rho_f \over \partial t} = -i \Bigl[ V_0 + \ol{V}_1, \rho_f \Bigr] -
2 \,
\GF^2 \Sc{A} \, \Bigl[(g^e)^2 \rho_f + \rho_f (g^e)^2  - 2 g^e \rho_f g^e
\Bigr],
\eeq
where $\rho_f$ is the $2\times 2$ neutrino-sector density matrix in flavour
space, $V_0
\approx k + {m^\dagger m \over 2k} + \cdots$, where $k$ is the neutrino
momentum
and $m$ is its mass matrix. $\ol{V}_1(t) \equiv \sqrt2 \, \GF \Avg{ g^e n_e(t)
+
g^n n_n(t)}$ is the first-order effective interaction, with $g^e = {\rm diag}
\,
(1,0)$, and $g^n = {\rm diag} \, (-\, \hf , - \, \hf )$ being the $2 \times 2$
matrices which describe the charged-current (neutral-current) neutrino coupling
to
the electron (neutron) density, $n_e$ ($n_n$). Because $g^n$ is proportional to
the unit matrix, the neutron density drops out of eq.~(\ref{explme}). The first
term of eq.~({\ref{explme}) gives the usual MSW evolution. All second-order
terms
are proportional to the coefficient:
\eq
\label{adefmatter}
\Sc{A}(t) =  \int_{t'}^t d\tau \; \Avg{\delta n_e(t) \; \delta n_e(\tau)} ,
\eeq
which controls the size of fluctuation effects.

It is instructive to compare the order of magnitude of the various terms in
eq.~(\ref{explme}). The flavour-changing part of $V_0$ is $m^\dagger m/k$;
while
the term involving $V_1$ is $\sim \GF g^e \Avg{n_e}$. These two terms are
comparable over part of the solar interior, and their interplay gives rise to
the
MSW resonant oscillation. The contribution of the remaining term is of order
$\GF^2 \Avg{\delta n_e \delta n_e} \tau_c$, which is competitive with the
others
if $\GF \Avg{\delta n_e \delta n_e} \tau_c \sim \Avg{n_e}$. Writing
$\Avg{\delta
n_e \delta n_e} = \epsilon^2 \Avg{n_e}^2$, this condition becomes $\epsilon^2
\GF
\Avg{n_e} \tau_c \sim 1$. Since, deep within the solar interior, $\GF \Avg{n_e}
\lsim (100 \; {\rm km})^{-1}$, significant fluctuation effects require
$\epsilon^2
\tau_c \gsim 100$ km.

These estimates are borne out by more detailed calculations \cite{ourpaper}.
When
applied to thermal fluctuations, they reproduce the standard result for
neutrino
scattering from thermally-distributed electrons, giving a negligible scattering
rate within the sun. We have identified only one potential source of
fluctuations
within the sun whose effects may be nonnegligible. These fluctuations arise due
to
the variation, over space and time, of the mean electron density within the
sun.
The propagation of any particular neutrino through such variations is perfectly
well described by the MSW treatment. In general, however, successive neutrinos
do
{\it not} encounter the {\it same} electron density profile. This could be
because
different neutrinos are produced at different places within the sun, and so
pass
through different regions while {\it en route} to the earth. Alternatively,
successive neutrinos produced at the same point within the sun could pass
through
different electron densities because the density profile itself generally
varies
in time -- such as when helioseismic waves are present. As a result, the
neutrino
flux to which a detector is exposed can be thought to have experienced an {\it
ensemble} of density profiles, which must be averaged to obtain the integrated
neutrino signal as seen by a detector on earth.

We have computed \cite{ourpaper} how the MSW resonance is altered due to
two kinds of ensembles of this type which are of particular interest. These
ensembles differ in the nature of the correlations which they assume. One
choice
assumes fluctuations which are uncorrelated in position space, \ie\ $\delta
n_e(x)$ is uncorrelated with $\delta n_e(y)$ for sufficiently large $|\bf{x} -
\bf{y}|$. This kind of local fluctuation is typical of many condensed-matter
applications, and could arise in the sun from a number of sources. The second
choice consists of fluctuations which are uncorrelated in {\it momentum} space,
with $\delta n_e(k)$ uncorrelated with $\delta n_e(k')$, where $k \neq k'$
label an appropriate set of density-oscillation modes. This second kind of
ensemble is intended as a simple model of how neutrinos interact with
helioseismic
waves, since these have been observed to exist in the sun.

It is straightforward to generalize the MSW analysis to arrive at a variation
of
Parke's formula \cite{Parke}:
\eq
\label{Parkeform}
P_e(t) = \hf + \left( \hf - P_\ss{J} \right) \lambda \, \cos 2\theta_m(t')
\, \cos 2\theta_m(t) ,
\eeq
in which $P_\ss{J} = \exp\left[ - \, {\pi \over 2} \; \left( {\sin^2 2
\theta_\ss{V} \over \cos 2 \theta_\ss{V}} \right) \left( {\delta m^2 \, h \over
2
k} \right) \right] $ is the usual `jump' probability as one passes through the
resonance point, with $h$ is the scale height for the electron density and
$\theta_{\ss{V}}$, $\theta_{m}$ and $\delta m^2 = m_h^2 - m_\ell^2$ are the
vacuum and matter mixing angles, and the difference between the squares of the
masses, for the two neutrino states. Fluctuation effects enter
eq.~({\ref{Parkeform}) through the damping factor $\lambda = \exp\left[ - 2
\int_{t'}^t dx \; \GF^2 \Sc{A}(x) \sin^2 2\theta_m(x) \right]$.

Notice that the fluctuations dominantly act to {\it damp} the neutrino
oscillations. This damping is a reflection of the conversion of the incoming
neutrino from a pure to a mixed state as it interacts with the fluctuations.
The
damping ruins the MSW resonance condition, thereby reducing the effectiveness
of
the flavour change as the neutrino passes through the resonant point. Our
numerical
results (presented in the figures) show that small fluctuations first become
noticeable where the resonant flavour change is the strongest. Since ${}^7$Be
neutrinos are suppressed by MSW oscillations in this way, a deviation from the
MSW
prediction for the strength of the ${}^7$Be neutrino flux could be evidence for
electron density fluctuations deep within the solar interior.

For sufficiently large times, eq.~(\ref{Parkeform}) has the universal
prediction:
$P_e \to \hf$. This limit is also seen in our numerical integrations, which
agree
well with eq.~(\ref{Parkeform}) throughout parameter space. This suggests a new
solution for the solar neutrino problem: an approximately energy-independent
suppression of the solar neutrino flux by a factor of 2 due to decoherence by
solar
fluctuations. Such a suppression would look much like large-angle vacuum
oscillations, even though it could arise through fluctuations with small vacuum
mixing angles.

The figures present a summary of the results of a numerical integration of
eq.~(\ref{explme}), using eq.~(\ref{adefmatter}) for density fluctuations
which are uncorrelated in either position and momentum space \cite{ourpaper}.
As is
seen from these figures, deviations from MSW predictions require fluctuations
whose
amplitude is, at the very least, a few percent.

\vspace{0.5cm}

\centerline{\epsfxsize=9.5cm\epsfbox[45 430 550 750]{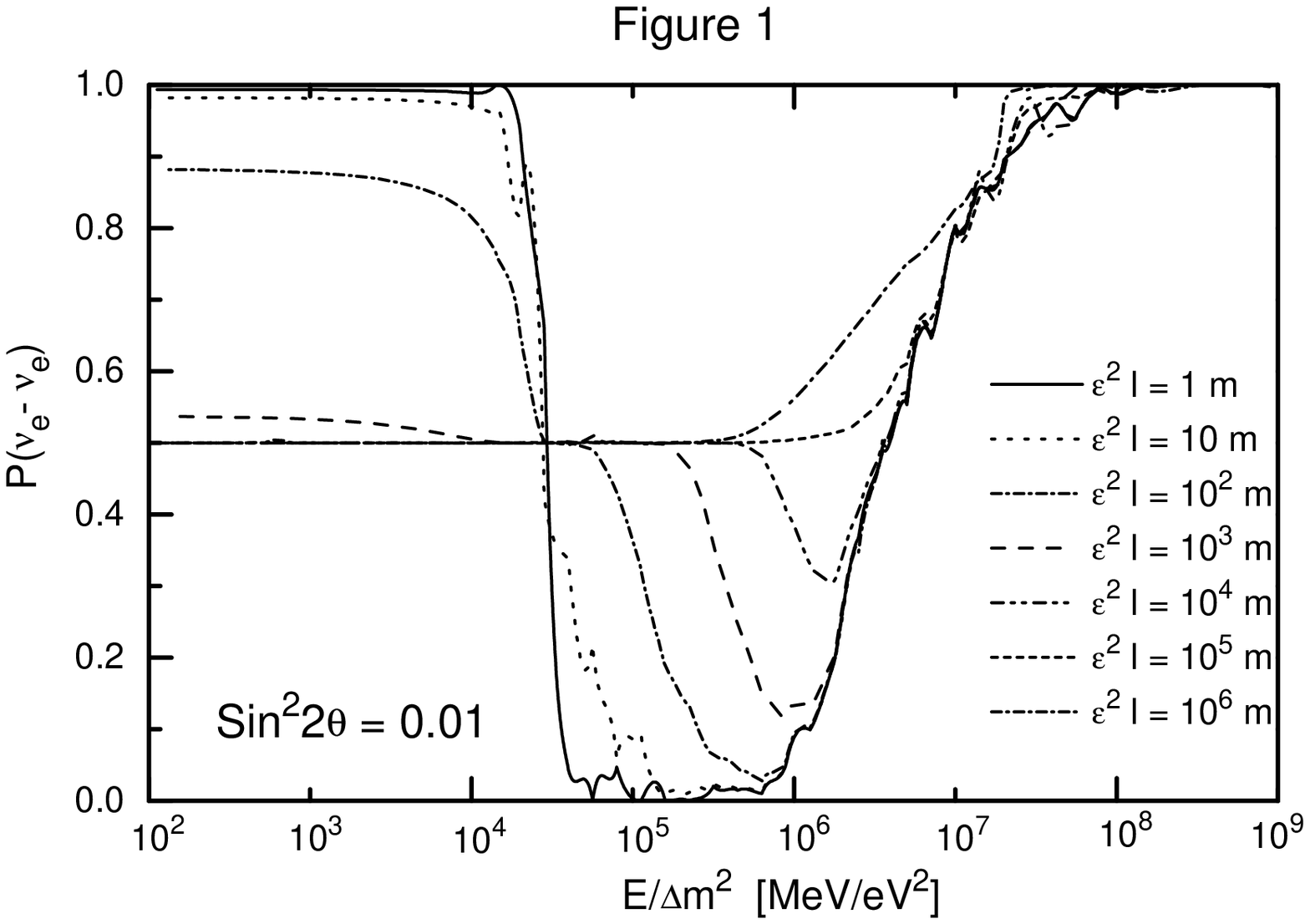}}
\bigskip\bigskip
\centerline{\epsfxsize=9.5cm\epsfbox[45 430 550 750]{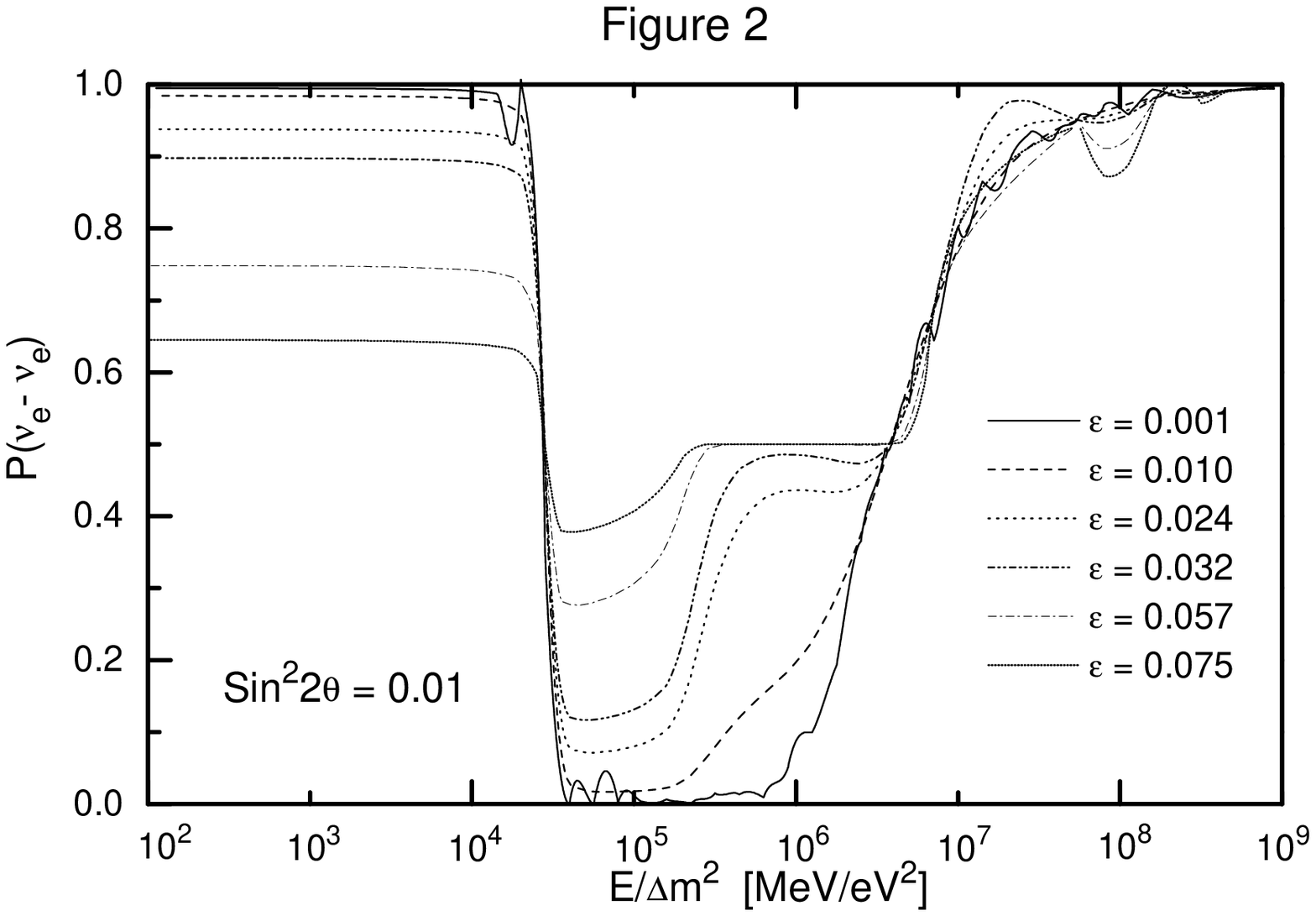}}

\begin{quote}
{\footnotesize The $\nu_e$ survival probability as a function of the neutrino
energy, in units of $\Delta m^2$. Figure 1 (Figure 2) uses fluctuations which
are uncorrelated in position (momentum) space.  The vacuum mixing angle
is taken to be 0.1. $epsilon$ denotes the fractional amplitude of the
fluctuation
while in Figure 1 $\ell$ represents the correlation length. }
\end{quote}

\section*{Acknowledgements}

We would like to thank the Conference organizers for providing such a splendid
setting, and for their kind invitation to present our work. Our funds have been
provided by the Natural Sciences and Engineering Research Council of Canada and
les Fonds pour la Formation de Chercheurs et l'Aide \`a la R\'echerche du
Qu\'ebec.

\def\pr#1{\it Phys.~Rev.~{\bf #1}}
\def\np#1{\it Nucl.~Phys.~{\bf #1}}
\def\pl#1{\it Phys.~Lett.~{\bf #1}}
\def\prc#1#2#3{{\it Phys.~Rev.~}{\bf C#1} (19#2) #3}
\def\prd#1#2#3{{\it Phys.~Rev.~}{\bf D#1} (19#2) #3}
\def\prl#1#2#3{{\it Phys. Rev. Lett.} {\bf #1} (19#2) #3}
\def\plb#1#2#3{{\it Phys. Lett.} {\bf B#1} (19#2) #3}
\def\npb#1#2#3{{\it Nuc. Phys.} {\bf B#1} (19#2) #3}
\def\etal{{\it et.al. \/}}

\pagebreak

\bibliographystyle{unsrt}

\end{document}